\documentclass{elsarticle}

\usepackage{graphicx,xcolor}
\usepackage{times}
\usepackage{amsmath,amssymb}
\usepackage[colorlinks,citecolor=red]{hyperref}

\newcommand\ket[1]{\mathinner{\lvert{\textstyle#1}\rangle}}
\newcommand{\Braket}[1]{\mathinner{\left\langle{#1}\right\rangle}}

\newcommand{\comm}[2]{[#1,#2]}
\newcommand{\acomm}[2]{\{#1,#2\}}

\newcommand{\bfk}{{\mathbf{k}}}

\newcommand{\re}{{\rm Re}}
\newcommand{\im}{{\rm Im}}

\newcommand{\Ni}{{N$_1$\ }}
\newcommand{\Nii}{{N$_2$\ }}

\newcommand{\eqnref}[1]{Eq.~(\ref{#1})}
\newcommand{\eqnsref}[1]{Eqs.~(\ref{#1})}

\newcommand{\figref}[1]{Fig.~\ref{#1}}
\newcommand{\figsref}[1]{Figs.~\ref{#1}}
\newcommand{\Figref}[1]{Figure~\ref{#1}}

\newcommand{\secref}[1]{Sec.~\ref{#1}}
\newcommand{\secsref}[1]{Secs.~\ref{#1}}
\newcommand{\Secref}[1]{Section~\ref{#1}}

\begin{document}

\title{Effects of Majorana bound states on dissipation and charging in a
  quantum resistor-capacitor circuit}

\author{Minchul Lee}

\address{
  Department of Applied Physics and Institute of Natural Science, College of Applied Science, Kyung Hee University, Yongin 17104, Korea
}

\ead{minchul.lee@khu.ac.kr}

\begin{abstract}
  We investigate the effects of Majorana bound states on the ac response of a
  quantum resistor-capacitor circuit which is composed of a topological
  superconducting wire whose two ends are tunnel-coupled to a lead and a
  spinless quantum dot, respectively. The Majorana states formed at the two
  ends of the wire are found to suppress completely or enhance greatly the
  dissipation, depending on the strength of the overlap between two Majorana
  modes and/or the dot level. We compare the relaxation resistance and the
  quantum capacitance of the system with those of non-Majorana counterparts to
  find that the effects of the Majorana state on the ac response are genuine
  and cannot be reproduced in ordinary fermonic systems.
\end{abstract}

\maketitle

\section{Introduction}
\label{sec:introduction}

Topological superconductors \cite{Hasan2010nov} has drawn a great interest of
solid-state physics society for last decade because they can host quasiparticle
excitations behaving like the Majorana fermions that are their own
anti-particles
\cite{Kitaev2001oct,Alicea2012jul,Leijnse2012dec,Beenakker2013apr,Stanescu2013jun,Elliott2015feb}. Due
to the inherent topological protection, Majorana quasiparticles, being formed
at the edges of the topological superconductors, can behave as nonlocal qubits
being resistant to decoherence phenomena \cite{Kitaev2001oct}. In addition,
they can perform non-Abelian statistics, making them the fundamental basis for
the realization of topological quantum computation \cite{Nayak2008sep}. Since
the realization of the elusive particles in solid-state systems was proposed,
many experimental implementations for the Majorana systems have been
reported. Most of them were based on a nanowire with strong spin-orbit
interaction put in proximity to a $s$-wave superconductor and exposed to a
magnetic field \cite{Mourik2012may,Das2012nov,Deng2012nov,Churchill2013jun,%
  Finck2013mar,Rokhinson2012sep,Albrecht2016mar}, where the evidence for
Majorana states were given by the appearance of the zero-bias anomaly in tunnel
spectroscopy or the fractional ac Josephson effect \cite{Jiang2011nov}.
Another experimental setup used a magnetically ordered atomic chain on a
superconducting surface and the spatially-resolved spectroscopy revealed the
existence of zero energy states at its ends
\cite{NadjPerge2014oct,Feldman2016nov}. While the unambiguous detection of the
Majorana state is still questioned, it is time to implement Majorana-based
circuits from the topological superconductors and to investigate their
operation in the presence of the interaction with environment.

A recent study \cite{LeeMC2014aug} investigated the quantum resistor-capacitor
(RC) circuit in which a quantum dot is coupled to chiral Majorana modes formed
around the edge of a two-dimensional topological superconductor. The Majorana
modes are found to open a dissipative channel inside the dissipationless
superconductor and to be able to enhance or completely suppress the relaxation
resistance in a non-trivial way, distinguishing it from normal fermionic
channels \cite{Buttiker1993sep,Buttiker1993jun}.
One can expect that the similar effects on the dissipation may be observed in
topological superconducting wires (TSWs) which have localized Majorana bound
states at their ends. Unlike the two-dimensional case, however, the discrete
Majorana bound states alone cannot dissipate, so an additional dissipative
channel is required to form a RC circuit.

In this paper we investigate the ac response of a quantum RC circuit in which a
TSW is inserted between a spinless quantum dot and a spinless lead [see
\figref{fig:system}], where the latter undertakes the dissipation. We
scrutinize the effect of the Majorana bound states on the dissipation and the
charging by analyzing relaxation resistance and the quantum capacitance of the
RC circuit.
We have found that when two localized Majorana modes do not overlap, no
dissipation takes place. The vanishing relaxation resistance is attributed to
the exact cancellation between charge-conserving and Cooper-pair tunneling
processes. When the overlap is finite, the relaxation resistance becomes finite
and can be greatly enhanced at the resonance condition of the quantum dot.
For comparison, the same analysis is applied to non-Majorana counterparts, and
we have found that the dependence of the relaxation resistance and the quantum
capacitance on temperature and the values of parameters is clearly different
between the Majorana and non-Majorana systems.

This paper is organized as follows. \Secref{sec:system} is devoted to the
introduction of the model Hamiltonian and the derivation of the formulas for
the relaxation resistance and the quantum capacitance. In \secsref{sec:em0} and
\ref{sec:em}, we analyze them for the cases of zero and finite overlaps between
two Majorana bound states, respectively, and compare with those of non-Majorana
systems. We make a conclusion in \secref{set:conclusion}, summarizing our findings.

\section{Quantum RC Circuit and Admittance}
\label{sec:system}

\subsection{Model}

\begin{figure}[!t]
  \centering
  \includegraphics[width=.5\textwidth]{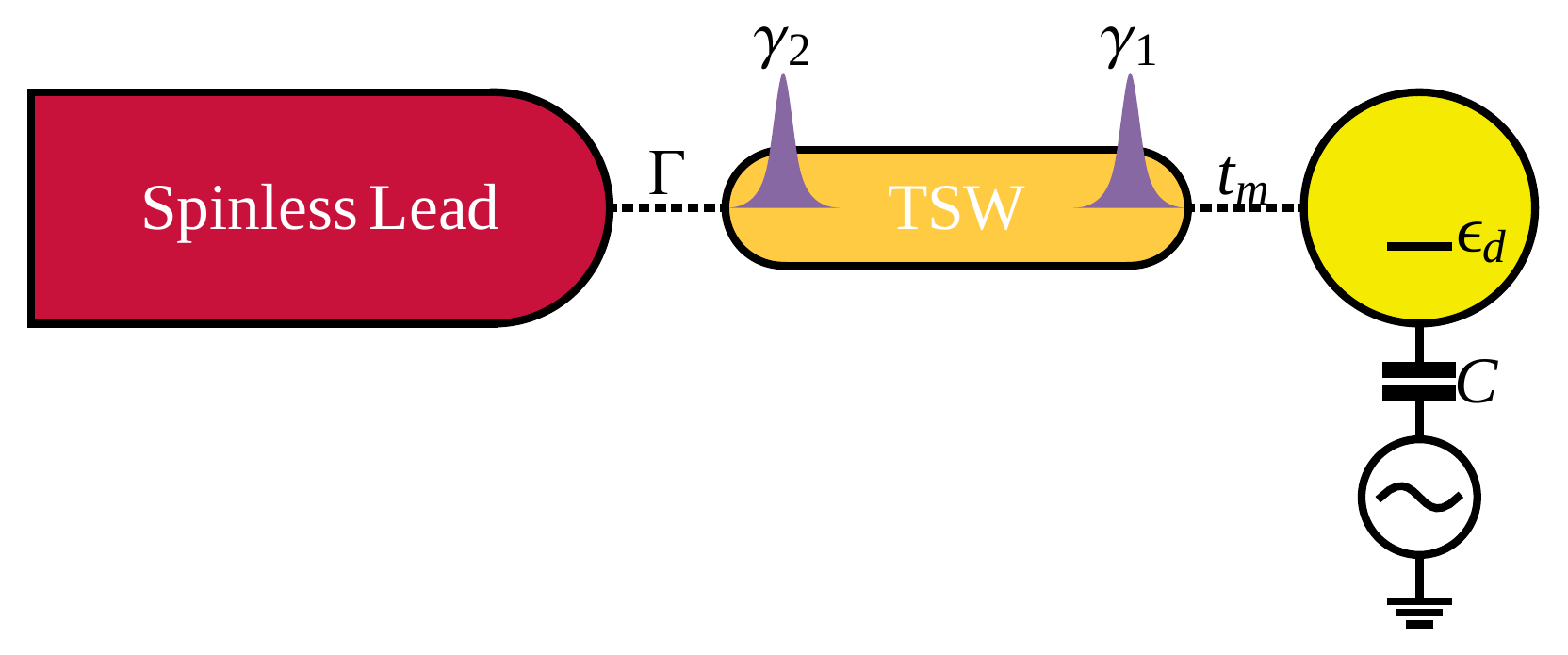}
  \caption{System configuration for a topological superconducting wire coupled
    to a lead and a quantum dot. Refer the definition of the symbols to the
    text. The Majorana bound states $\gamma_{1/2}$ are formed at two ends of
    the topological superconducting wire and each of them is tunnel-coupled to
    the quantum dot and the spinless lead, respectively.}
  \label{fig:system}
\end{figure}

\Figref{fig:system} shows the schematic configuration of the system of our
interest, in which two ends of a topological superconducting wire are
tunnel-coupled to a spinless quantum dot and a lead, respectively. In this
study we are interested in the low-energy physics below the bulk gap of the
superconducting wire, so the Hamiltonian for the TSW is described by the two
Majorana modes $\gamma_i = \gamma_i^\dag$ ($i=1,2$) localized at the ends as
\begin{equation}
  \label{eq:H:TSW}
  H_{\rm M}
  = 2i\epsilon_m \gamma_1 \gamma_2
  = 2\epsilon_m (f^\dag f - 1/2),
\end{equation}
where $\epsilon_m$ is the overlap between the Majorana modes and
$f = (\gamma_1 + i\gamma_2)/\sqrt2$ is the nonlocal fermion operator composed
of two Majorana operators. One end of the TSW is tunnel-coupled to a spinless
quantum dot, which is spin-polarized due to a sufficiently strong magnetic
field which is required to induce the topologically superconducting state in
the semiconductor wire. The quantum dot with a single spinless level
$\epsilon_d$ and its coupling to the TSW are described by
\begin{subequations}
  \label{eq:H:QD}
  \begin{align}
    H_{\rm QD}
    & = \left[\epsilon_d + e(U(t) - V(t))\right] n_d
    \\
    H_{\rm QD-M}
    & = t_m (d^\dag \gamma_1 + \gamma_1 d)
    = \frac{t_m}{\sqrt2} (d^\dag f + d^\dag f^\dag + (h.c.)),
  \end{align}
\end{subequations}
respectively. Here $n_d = d^\dag d$ is the occupancy operator, and the ac
voltage $V(t)$ upon the gate coupled to the quantum dot via a geometrical
capacitance $C$ induces the polarization charge on the dot and eventually the
internal potential $U(t)$. The latter is to be determined self-consistently
under the charge conservation condition. The dot level is coupled to only one
of the Majorana mode $\gamma_1$ with a tunneling strength $t_m$.
The second Majorana mode $\gamma_2$ at the other end of the TSW is coupled to
the lead. Since the Majorana mode has only a single spin component, only one of
two spin channels in the lead is coupled to the TSW. Hence, like the quantum
dot, the lead is considered to be spinless:
\begin{subequations}
  \label{eq:H:L}
  \begin{align}
    H_{\rm L}
    & = \sum_\bfk \epsilon_\bfk c_\bfk^\dag c_\bfk
    \\
    H_{\rm L-M}
    & = \sum_k t (c_\bfk^\dag \gamma_2 + \gamma_2 c_\bfk)
    =
    \sum_\bfk \frac{t}{\sqrt2} \left(i f^\dag c_\bfk - i f^\dag c_\bfk^\dag + (h.c.)\right),
  \end{align}
\end{subequations}
where the spinless conduction-electron operator $c_\bfk$ for momentum $\bfk$
defines a dispersion $\epsilon_\bfk$ and is coupled to $\gamma_2$ via the
tunneling amplitude $t$, which is assumed to be independent of momentum and
energy, for simplicity. The tunneling induces the hybridizations
$\Gamma = \pi\rho|t|^2$ between the TSW and the lead, where $\rho$ is the
density of states at the Fermi level in the lead.

\subsection{Relaxation Resistance and Quantum Capacitance}

We regard our system as a RC circuit
\cite{LeeMC2014aug,Buttiker1993sep,Buttiker1993jun,Nigg2006nov,LeeMC2011may,KhimH2013mar}
with respect to a weak time-dependent external gate voltage
\begin{math}
  V(t) = V_{\rm ac} \cos \omega t
\end{math}
applied on the quantum dot. The ac voltage $V(t)$, in the mean-field
approximation, induces the polarization charges $N(t)$ between the dot and the
gate, which in turn leads to the time-dependent potential $U(t) = |e|N(t)/C$
inside the dot.
Consequently, the applied voltage not only generates a current $I(t)$ between
the TSW and the dot, but also induces a displacement current
$I_d(t) = e (dN/dt) = - C (dU/dt)$ between the gate and the dot. The relation
between two currents is set by the charge conservation: $I(t) + I_d(t) = 0$.
Assuming that the gate-invariant perturbation, $V(t) - U(t)$, is sufficiently
small, the linear response theory leads to the relation,
$I(\omega) = g(\omega) (V(\omega) - U(\omega))$, where the
$g(t) = (ie/\hbar) \Braket{\comm{I(t)}{n_d}}\Theta(t)$ is the equilibrium
correlation function between the occupation operator $n_d$ and the current
operator $I = e (dn_d/dt)$.  Then, the dot-lead impedance
$Z(\omega) = V(\omega)/I(\omega)$, which is experimentally accessible, can be
expressed as $Z(\omega) = 1/(-i\omega C) + 1/g(\omega)$, by using the
self-consistent condition, $I(\omega) = - I_d(\omega) = -i\omega CU(\omega)$.
The quantum correction to the impedance then gives rise to the relaxation
resistance and the quantum correction to the capacitance:
\begin{equation}
  R_q(\omega)
  = \re\left[\frac{1}{g(\omega)}\right]
  \quad\text{and}\quad
  C_q(\omega)
  = \im\left[\frac{\omega}{g(\omega)}\right]^{-1}.
\end{equation}

In order to calculate the admittance, we adopt the Wingreen-Meir formalism
\cite{Wingreen1993sep,Jauho1994aug} which derives directly the current formula
for arbitrary gauge-invariant perturbation $V_g(t)-U(t)$ and then obtains the
admittance by considering the linear response only. We refer the details of the
derivation to Ref.~\cite{LeeMC2014aug}. The system studied in
Ref.~\cite{LeeMC2014aug}, being different from ours, shares the key features:
(1) The ac voltage is applied to the spinless quantum dot and (2) the dot is
tunnel-coupled to a single reservoir (directly or indirectly) with no other
channel for dissipation. Therefore, the general expression of the admittance in
terms of the dot Green's function, derived in Ref.~\cite{LeeMC2014aug}, applies
to our case as well:
\begin{align}
  \label{eq:g}
  \begin{split}
    g(\omega)
    & =
    \frac{\omega}{R_Q} \!\!
    \int\! d\omega' f(\omega')
    \left[
      G_d^R(\omega'-\omega) \sigma_3 \{G_d^R(\omega')-G_d^A(\omega')\}
    \right.
    \\
    & \qquad\qquad\qquad\qquad\left.\mbox{}
      +
      \{G_d^R(\omega')-G_d^A(\omega')\} \sigma_3 G_d^A(\omega'+\omega)
    \right]_{11},
  \end{split}
\end{align}
where $R_Q \equiv h/e^2$, $f(\epsilon)$ is the Fermi distribution function at
temperature $T$, and $\sigma_i$ ($i=0,1,2,3$) are the Pauli matrices in Nambu
space. The dot Green's function over the Nambu space is defined by
\begin{equation}
  G_d^{R/A}(t-t')
  =
  \mp i\Theta(\pm(t-t'))
  \begin{bmatrix}
    \Braket{|\acomm{d(t)}{d^\dag(t')}|} & \Braket{|\acomm{d(t)}{d(t')}|}
    \\
    \Braket{|\acomm{d^\dag(t)}{d^\dag(t')}|}
    & \Braket{|\acomm{d^\dag(t)}{d(t')}|}
  \end{bmatrix}.
\end{equation}

Since our system is effectively non-interacting, it is quite straightforward to
calculate the dot Green's function, which is given by
\begin{equation}
  \label{eq:G}
  G_d^R(\omega)
  = [G_d^A(\omega)]^\dag
  = \left[g_d^R(\omega) - \Sigma^R(\omega)\right]^{-1}
\end{equation}
with the unperturbed dot Green's function
$g_d^R(\omega) = (\omega + i\eta - \sigma_3\epsilon_d/\hbar)^{-1}$ and the
self-energy
\begin{equation}
  \label{eq:selfenergy}
  \Sigma^R(\omega)
  =
  \frac{|t_m|^2}{\hbar^2}
  \frac{\omega + 2i\Gamma}{(\omega+i\eta)(\omega+2i\Gamma) - (2\epsilon_m/\hbar)^2}
  (\sigma_0 - \sigma_1),
\end{equation}
where $\eta$ is positive infinitesimal number.  As expected, the Green's
function has finite off-diagonal components which reflects the presence of the
superconductivity.

In our study, in order to distinguish the physical features peculiar to
Majorana fermions from non-Majorana ones, we consider non-Majorana counterpart
systems in which the Majorana bound modes are replaced by an ordinary fermionic
mode. Specifically, two systems, N$_i$ ($i=1,2$) are examined: In the \Ni
system, the dot and the lead is connected via a single (local) fermionic level
$f$ in the wire, whose level energy is now given by $2\epsilon_m$ [see
\eqnref{eq:H:TSW}]. This system is implemented by eliminating the abnormal
terms such as $d^\dag f^\dag$ and $c_\bfk^\dag f^\dag$ in both $H_{\rm QD-M}$
and $H_{\rm L-M}$, which turns off the superconductivity in the wire. The \Nii
system further assumes that the localized bound state $f$ is coupled only to
the dot, while being disconnected from the lead. The two systems are
implemented by using the self-energies
\begin{equation}
  \label{eq:selfenergy:0}
  \Sigma^R_{\rm N1}(\omega)
  =
  \frac{|t_m|^2/\hbar^2}{\omega - 2\epsilon_m\sigma_3/\hbar + i\Gamma/2}
  \quad\text{and}\quad
  \Sigma^R_{\rm N2}(\omega)
  =
  \frac{|t_m|^2/\hbar^2}{\omega + i \eta - 2\epsilon_m\sigma_3/\hbar},
\end{equation}
respectively, in the dot Green's function, \eqnref{eq:G}.

\subsection{Excitation Spectrum of Dot-Wire Subsystem}

\begin{figure}[!t]
  \centering
  \includegraphics[width=.5\textwidth]{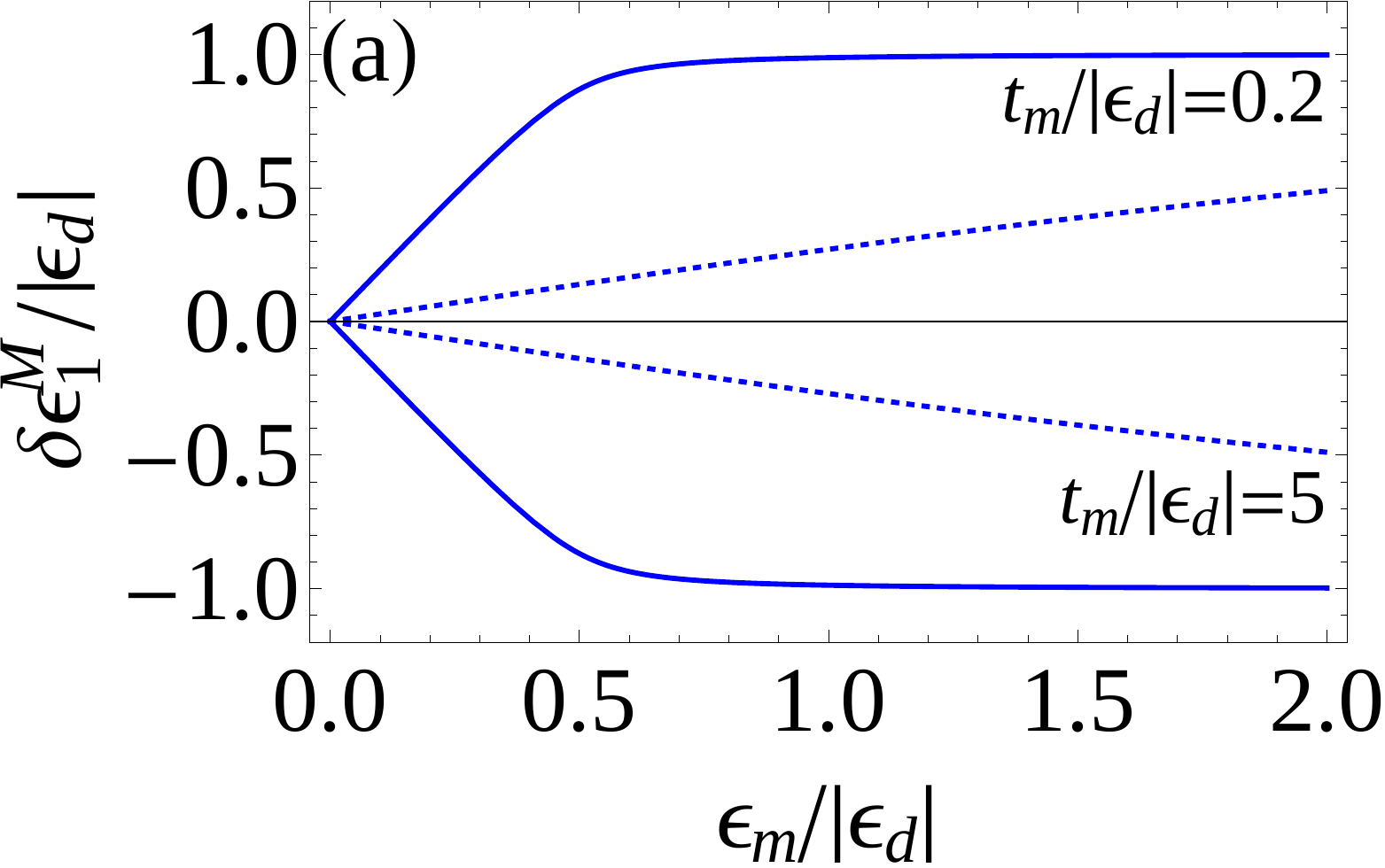}%
  \includegraphics[width=.5\textwidth]{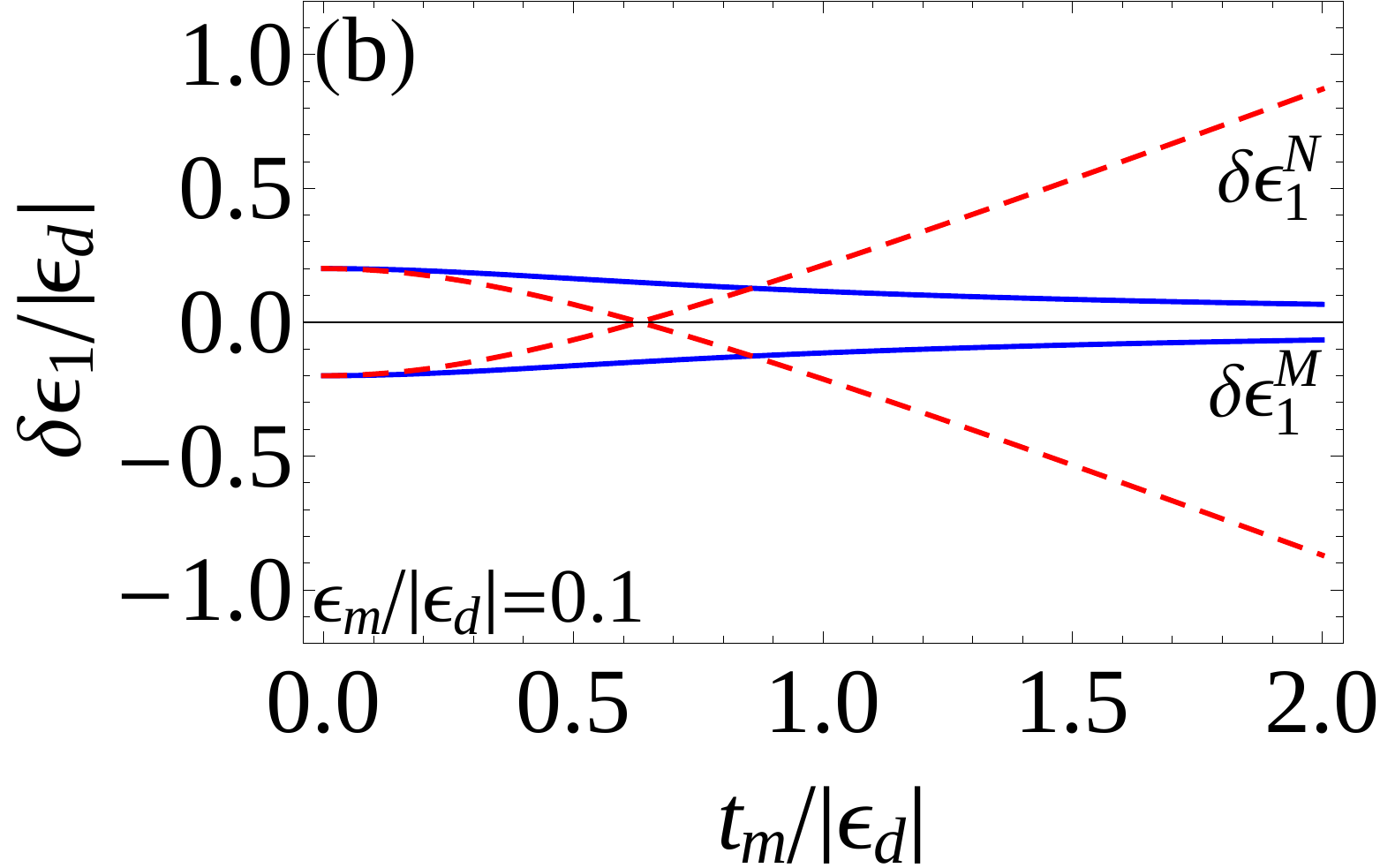}%
  \caption{(a) Excitation energies $\pm\delta\epsilon_1^M$ as functions of
    $\epsilon_m$ for two different values of $t_m$ and (b)
    $\pm\delta\epsilon_1^M$ (solid line) and $\pm\delta\epsilon_1^N$
    (dashed line) as functions of $t_m$ at a fixed value of $\epsilon_m$.}
  \label{fig:exc}
\end{figure}

For later use, we diagonalize the dot-wire subsystem disconnected from the lead
and obtain the excitation spectrum for our system and the N$_{1/2}$ system. By
diagonalizing $H_{\rm QD} + H_{\rm QD-M} + H_{\rm M}$, one obtains four
eigenenergies, $\epsilon_{1\pm}$ and $\epsilon_{2\pm}$ defined as
$\epsilon_{1\pm}^M \equiv
(\epsilon_d\pm\sqrt{(\epsilon_d-2\epsilon_m)^2+2|t_m|^2})/2$
and
$\epsilon_{2\pm}^M \equiv
(\epsilon_d\pm\sqrt{(\epsilon_d+2\epsilon_m)^2+2|t_m|^2})/2$
for the Majorana system, and $\epsilon_{1\pm}^N = \epsilon_{1\pm}^M$,
$\epsilon_{2+}^N = \epsilon_d+\epsilon_m$, and $\epsilon_{2-}^N = -\epsilon_m$
for the N$_{1/2}$ system. The eigenstates for the eigenenergies
$\epsilon_{1\pm}$ are built from the linear combinations of $d^\dag\ket0$ and
$f^\dag\ket0$, while those for $\epsilon_{2\pm}$ are from $\ket0$ and
$d^\dag f^\dag \ket0$. Hence the excitation energies with respect to the
$d$-particle are
\begin{align}
  \label{eq:excitation}
  \pm\delta\epsilon_1=\epsilon_{2\pm} - \epsilon_{1\pm}
  \quad\text{and}\quad
  \pm\delta\epsilon_2=\epsilon_{2\pm} - \epsilon_{1\mp}.
\end{align}
Typical plots of $\delta\epsilon_1^{M/N}$ are drawn in \figref{fig:exc} for
later use.

\section{Relaxation Resistance and Quantum Capacitance: $\epsilon_m=0$ Case}
\label{sec:em0}

First, we investigate the case in which the two Majorana modes have no overlap
between them, that is, $\epsilon_m=0$. It corresponds to the ideal case for
Majorana braiding and Majorana-based quantum computation because no
time-dependent phase change of Majorana states occurs. In this case, the self
energy, \eqnref{eq:selfenergy}, is quite simplified to
\begin{equation}
  \label{eq:selfenergy:em0}
  \Sigma^R(\omega)
  = \frac{|t_m|^2}{\hbar^2} \frac{1}{\omega+i\eta} (\sigma_0 - \sigma_1).
\end{equation}
The key characteristics of the above self energy is that the effect of the lead
hybridization disappears completely: it has no dependence on $\Gamma$. There
are two arguments to explain it. The first one is simple: The condition
$\epsilon_m=0$ breaks the coupling between two Majorana modes so that the
system is divided into two independent parts. The dot coupled to one of the
Majorana modes, therefore, is completely disconnected from the lead coupled to
another Majorana mode, leaving no dependence on $\Gamma$ in $G^R(\omega)$.
The second argument, instead, interprets it in terms of the fermionic mode $f$,
not the Majorana modes $\gamma_i$. As can be seen in \eqnsref{eq:H:QD} and
(\ref{eq:H:L}), the $f$-particle tunneling to both the dot and the lead is
possible, and the wire is simply on resonance for $\epsilon_m=0$. The dot, in
this view, is not apparently decoupled from the lead. However, the coupling of
$f$-particle to the dot and the wire involves the Cooper-pairing tunneling such
as $d f$ and $c_\bfk f$ as well as the particle-conserving one such as
$d^\dag f$ and $c_\bfk^\dag f$. The point is that these two tunnelings can
cancel out each other. This cancellation can be more easily understood in the
perturbation language. In the weak wire-lead tunneling limit, the perturbation
theory gives rise to correction terms related to two tunneling terms and their
weights are proportional to $1/(\epsilon_\bfk - \epsilon_m)$ and
$1/(-\epsilon_\bfk - \epsilon_m)$ for the charge-conserving ($c_\bfk^\dag f$)
and the Cooper-pair ($c_\bfk f$) tunnelings. For $\epsilon_m=0$, the two terms
are same in magnitude and have opposite signs, leading to complete
cancellation.
This vanishing effect is common in Majorana systems \cite{LeeMC2014aug},
because the Majorana modes, being anti-particles of themselves, have particle
and hole components in equal magnitude.

Using the self-energy, \eqnref{eq:selfenergy:em0} and the corresponding Green's
function, the expression for the admittance, \eqnref{eq:g} is given by
\begin{equation}
  g(\Omega)
  =
  i \frac{2\pi}{R_Q}
  \frac{|t_m|^2 \hbar\omega}{\epsilon_w (\epsilon_w^2 - (\hbar\omega)^2)}
  \tanh \frac{\beta\epsilon_w}{2}
\end{equation}
where $\epsilon_w \equiv \sqrt{\epsilon_d^2 + 2|t_m|^2}$ is the excitation
energy due to the hybridization of the dot and the Majorana mode and
$\beta \equiv 1/k_BT$. The admittance is purely imaginary, meaning that there
is no dissipation, $R_q(\omega) = 0$ for all frequencies $\omega$, because no
dissipation channel is connected.
The quantum capacitance is then
\begin{align}
  \label{eq:Cq:em0}
  C_q(\omega)
  =
  -e^2 \frac{|t_m|^2}{\epsilon_w(\epsilon_w^2-(\hbar\omega)^2)}
  \tanh\frac{\beta\epsilon_w}{2}.
\end{align}
It should be noted that its temperature dependence is exponential, which
reflects the fact that the dot-TSW system defines sharp energy eigenstate with
no broadening.

Now we consider the non-Majorana systems and identify the features unique to the
Majorana systems.
The \Ni system which has no Cooper-paring tunneling opens a dissipation channel,
so $R_q(\omega)$ is finite and depends on the values of parameters. For
example, for $|t_m|\ll\Gamma$, the low-frequency resistance
$R_{q0}\equiv R_q(\omega\to0)$ is $R_Q/2$, the universal value of $R_q$ when
only a single dissipation channel is involved
\cite{Buttiker1993sep,Buttiker1993jun}.
In the \Nii system, the dot is still disconnected from the lead so as in the
Majorana case no dissipation happens: $R_q(\omega)=0$. However, the temperature
dependence of the quantum capacitance is different from that in the Majorana
case, \eqnref{eq:Cq:em0}:
\begin{align}
  \label{eq:Cq:N2:em}
  C_q(\omega)
  =
  -e^2 \frac{|t_m|^2}{\epsilon_w(\epsilon_w^2-(\hbar\omega)^2)}
  \frac{\tanh \frac{\beta\epsilon_w}{2}}%
  {1 + \cosh\frac{\beta\epsilon_d}{2}/\cosh\frac{\beta\epsilon_w}{2}}.
\end{align}
This difference can be understood by comparing the excitation spectra of the
Majorana and non-Majorana systems.
By diagonalizing the dot-wire subsystems, the excitation energies,
\eqnref{eq:excitation} are obtained as $\delta\epsilon_1=0$ (doubly
degenerate), $\delta\epsilon_2=\epsilon_w$ for the Majorana systems and
$\delta\epsilon_1=(\epsilon_d-\epsilon_w)/2$ and
$\delta\epsilon_2=(\epsilon_d+\epsilon_w)/2$ for the \Nii system. Since the
charging and discharging via the degenerate zero-energy excitation cancel out
each other, $C_q \propto f(\epsilon_w) - f(-\epsilon_w)$ for the Majorana
system, while
$C_q \propto f((\epsilon_d+\epsilon_w)/2)-f((\epsilon_d-\epsilon_w)/2)$ for the
\Nii system, resulting in \eqnsref{eq:Cq:em0} and (\ref{eq:Cq:N2:em}),
respectively. Therefore, the difference is attributed to the pinning to the
zero-energy excitation of the Majorana system.


\section{Relaxation Resistance and Quantum Capacitance: $\epsilon_m\ne0$ Case}
\label{sec:em}

\begin{figure}[!t]
  \centering
  \includegraphics[width=.5\textwidth]{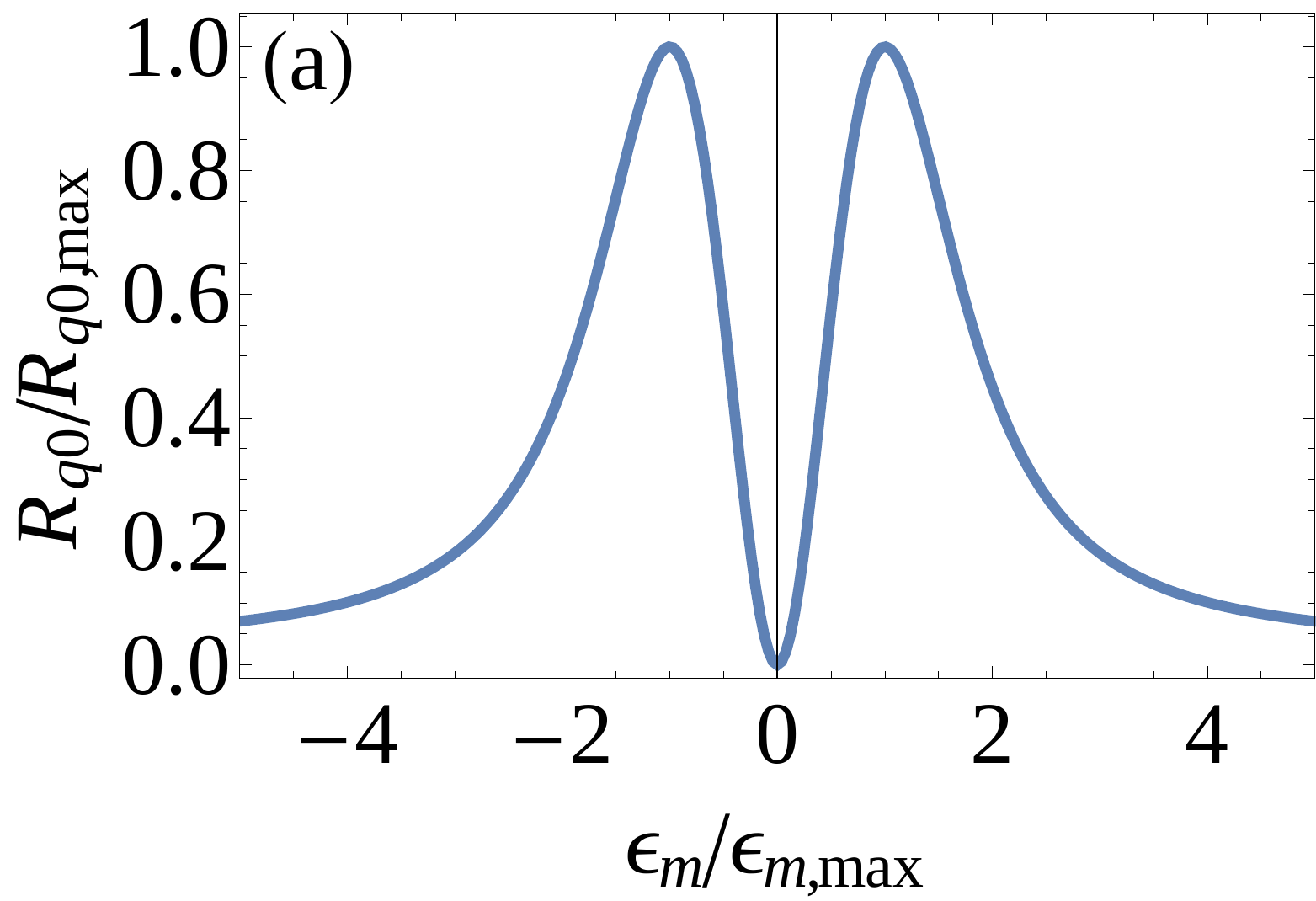}%
  \includegraphics[width=.5\textwidth]{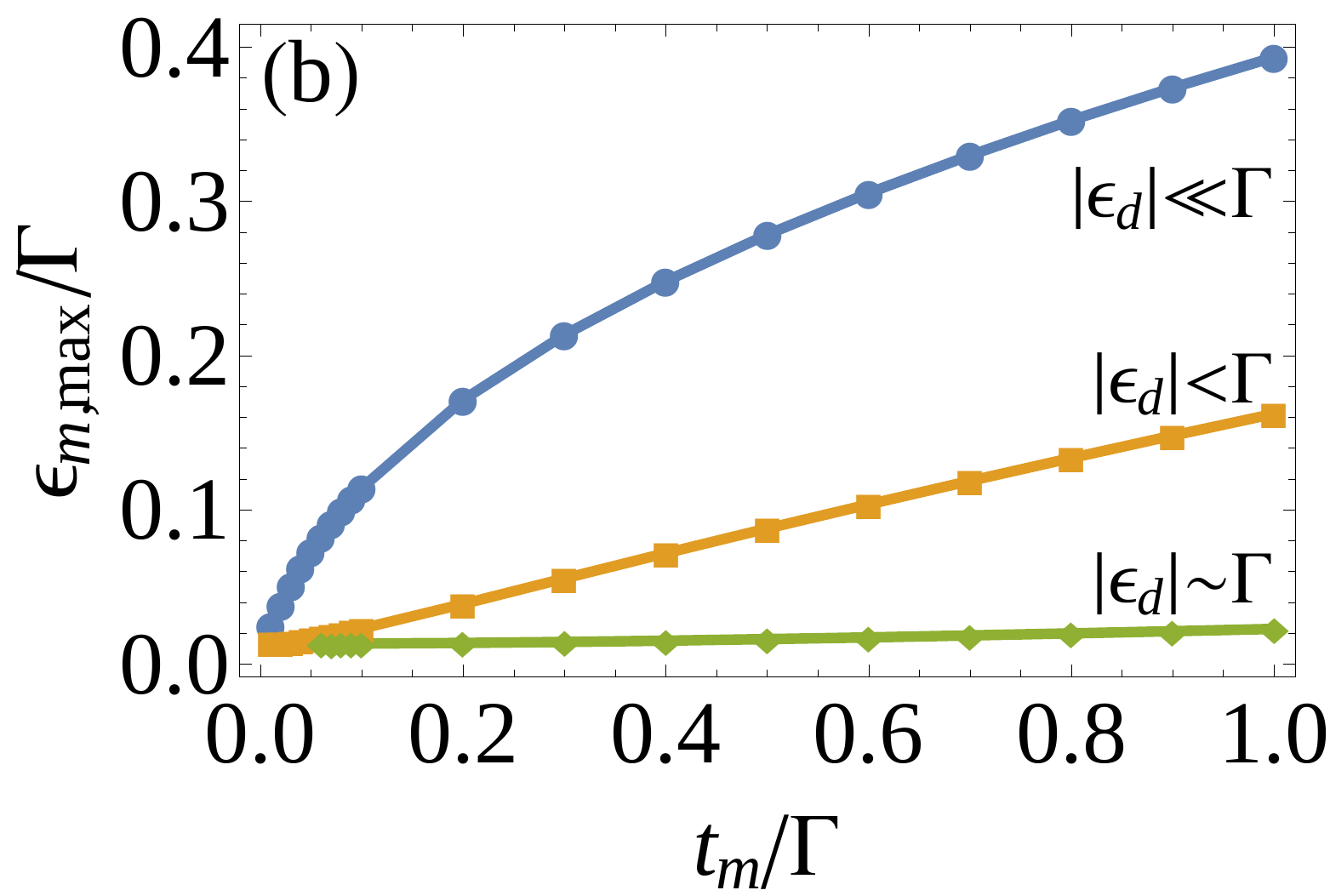}%

  \includegraphics[width=.5\textwidth]{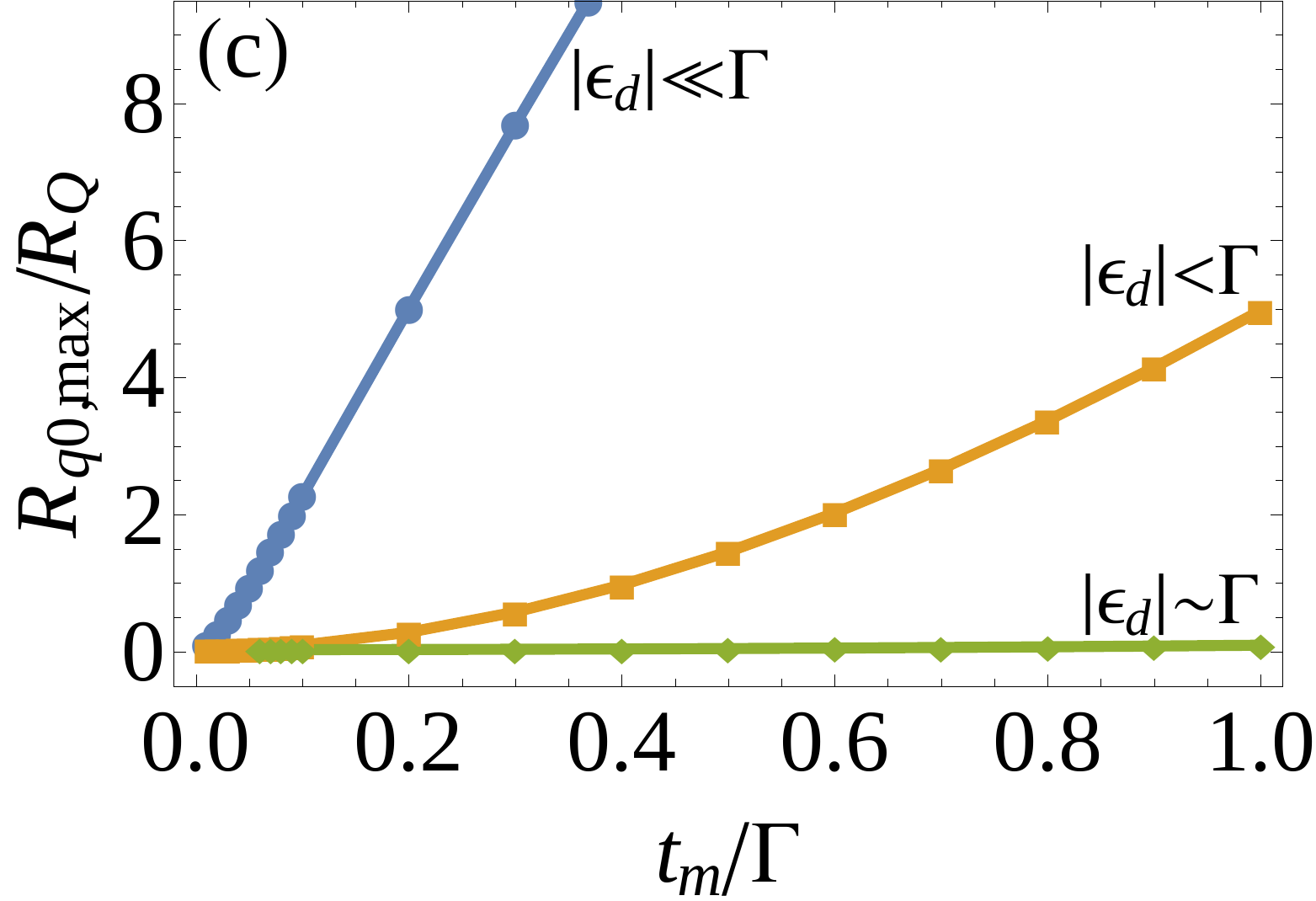}%
  \includegraphics[width=.5\textwidth]{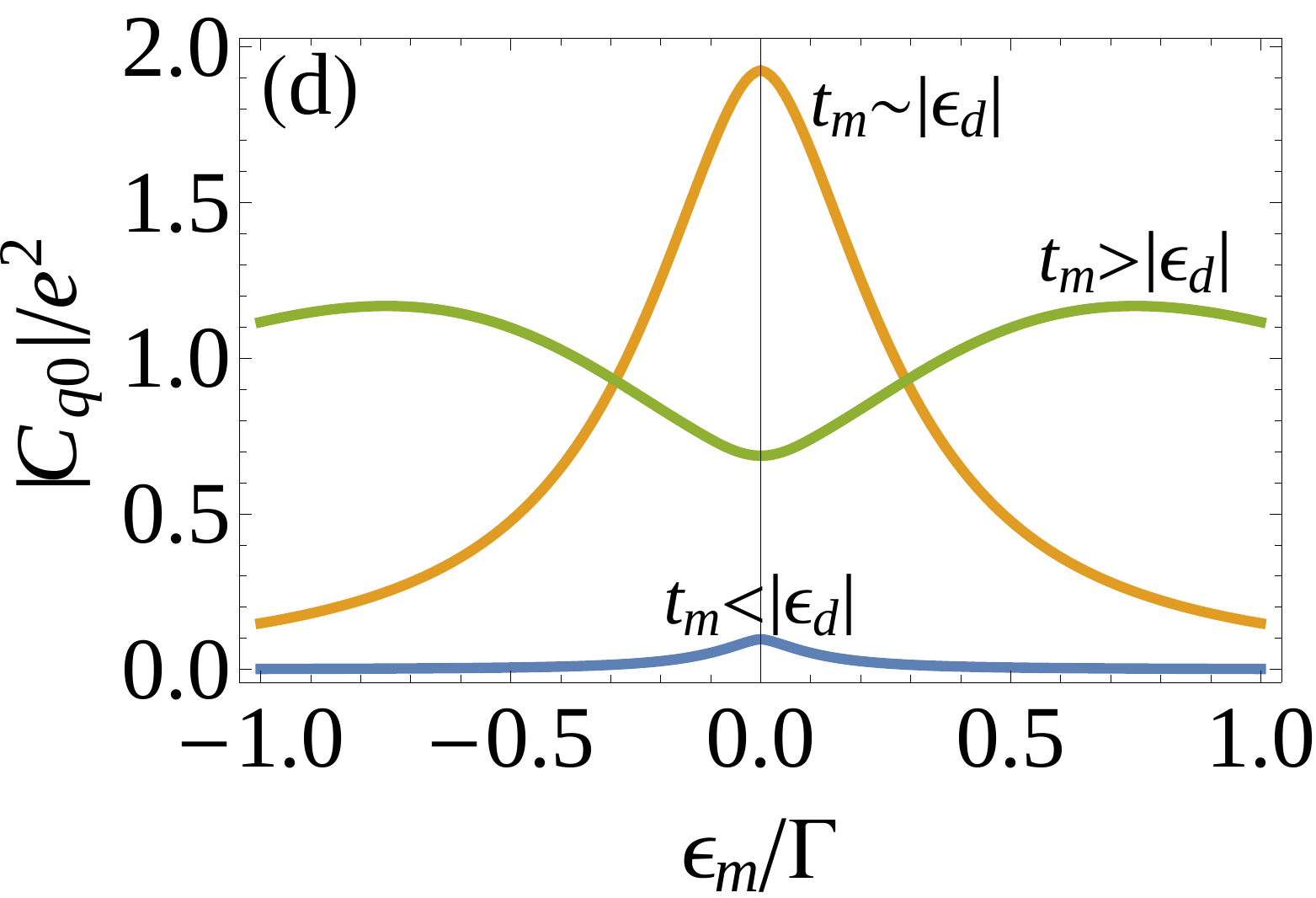}%
  \caption{(a) The low-frequency relaxation resistances $R_{q0}/R_{q0,\rm max}$
    as functions of $\epsilon_m/\epsilon_{m,\rm max}$ (see the text for the
    definition of $R_{q0,\rm max}$ and $\epsilon_{m,\rm max}$) at zero
    temperature, and (b,c) $\epsilon_{m,\rm max}$ and $R_{q0,\rm max}$ as
    functions of $t_m/\Gamma$ in different regimes: $|\epsilon_d|\ll \Gamma$,
    $|\epsilon_d|<\Gamma$, and $|\epsilon_d|\gtrsim\Gamma$. (d) The
    low-frequency quantum capacitances $C_{q0}$ as functions of
    $\epsilon_m/\Gamma$ at zero temperature in different regimes:
    $t_m<|\epsilon_d|$, $t_m\sim|\epsilon_d|$, and $t_m>|\epsilon_d|$ with
    $\epsilon_d/\Gamma = -0.1$.}
  \label{fig:RqCq}
\end{figure}

Finite $\epsilon_m$ connects the dot and the lead via the non-local Majorana
modes. In terms of $f$-particle tunnelings, the exact cancellation between the
charge-conserving and Cooper-pairing processes is lift for non-zero
$\epsilon_m$. \Figref{fig:RqCq} shows the dependence of $R_{q0}$ on the value
of $\epsilon_m$ at zero temperature. While $R_{q0}$ vanishes at $\epsilon_m=0$,
it becomes non-zero for finite values of $\epsilon_m$. \Figref{fig:RqCq}~(a)
clearly shows that $R_{q0}$ has two symmetric side peaks at
$\epsilon_m = \pm\epsilon_{m,\rm max}$ with the height $R_{q0,\rm max}$, where
the peak position $\epsilon_{m,\rm max}$ and its height $R_{q0,\rm max}$ vary
with the values of $\epsilon_d/\Gamma$ and $t_m/\Gamma$, as shown in
\figsref{fig:RqCq}~(b) and (c). As explained in details in
Ref.~\cite{LeeMC2011may,LeeMC2014aug}, the relaxation resistance measures the
dissipation due to the relaxation of the particle-hole (p-h) pairs generated
via the dot-lead tunneling. In the language of the perturbation, the weight of
the generated p-h pairs depends on the energy of the intermediate virtual
states. More specifically, it is proportional to
$1/(|\epsilon_\bfk| + |\epsilon_m|)$. Therefore, the more p-h pairs are
generated for the smaller $\epsilon_m$. However, as explained in the previous
section, the cancellation between the two types of tunnelings is pronounced for
small values of $\epsilon_m$. As a result, $R_{q0}$ exhibits a narrow dip
inside a central wider peak around $\epsilon_m=0$. The height $R_{q0,\rm max}$
of the side peak increases with $t_m$ and is maximized at resonance
$\epsilon_d = 0$, since the more p-h pairs are generated in these
conditions. The peak position $\epsilon_{m,\rm max}$ or the width of the dip
also shows qualitatively similar features since for larger $t_m$ and smaller
$|\epsilon_d|$ the effect of the Majorana coupling $\epsilon_m$ becomes
relatively smaller and the cancellation becomes stronger, widening the dip. It
should be noted that the enhancement of $R_{q0}$ on the dot resonance condition
is also observed in the RC circuit made of chiral Majorana edge modes
\cite{LeeMC2014aug}, indicating that this enhancement comes from the nature of
the Majorana fermions.

The dependence of the low-frequency quantum capacitance
$C_{q0}\equiv C_q(\omega\to0)$ on $\epsilon_m$ at zero temperature is displayed
in \figref{fig:RqCq}~(d). Finite $\epsilon_m$ lifts the degeneracy at the
otherwise zero-energy excitations, making $\delta\epsilon_1$ finite [see
\figref{fig:exc}~(a)]. It means that the dot density of states $\rho_F$ at the
Fermi level decreases with increasing $|\epsilon_m|$. Since the quantum
capacitance at zero temperature is, though not exactly, proportional to
$\rho_F$, $C_{q0}$ also decreases with increasing $|\epsilon_m|$, as shown in
\figref{fig:RqCq}~(d).
For $t_m>\epsilon_d$, on the other hand, the strong hybridization between the
dot and the wire makes $\delta\epsilon_1$, though being finite, remain small
even for large $\epsilon_m$ [see \figref{fig:exc}~(a)]. So, the rapid decrease
of $C_{q0}$ with increasing $|\epsilon_m|$ is alleviated. In addition, as shown
in \figref{fig:RqCq}~(d), a weak side-peak structure appears, which is
attributed to the lead-driven broadening and the shift of the spectral weights.

We have found that the temperature dependence of $R_{q0}$ and $C_{q0}$ follows
the Fermi-liquid-like behavior: The correction due to small thermal
fluctuations is found to be proportional to $T^2$ as predicted by the
Sommerfeld expansion. One exception is when the dot is exactly at resonance
which induces the $\delta$-peak in the dot density of states, resulting in
exponential dependence.

\begin{figure}[!t]
  \centering
  \includegraphics[width=.5\textwidth]{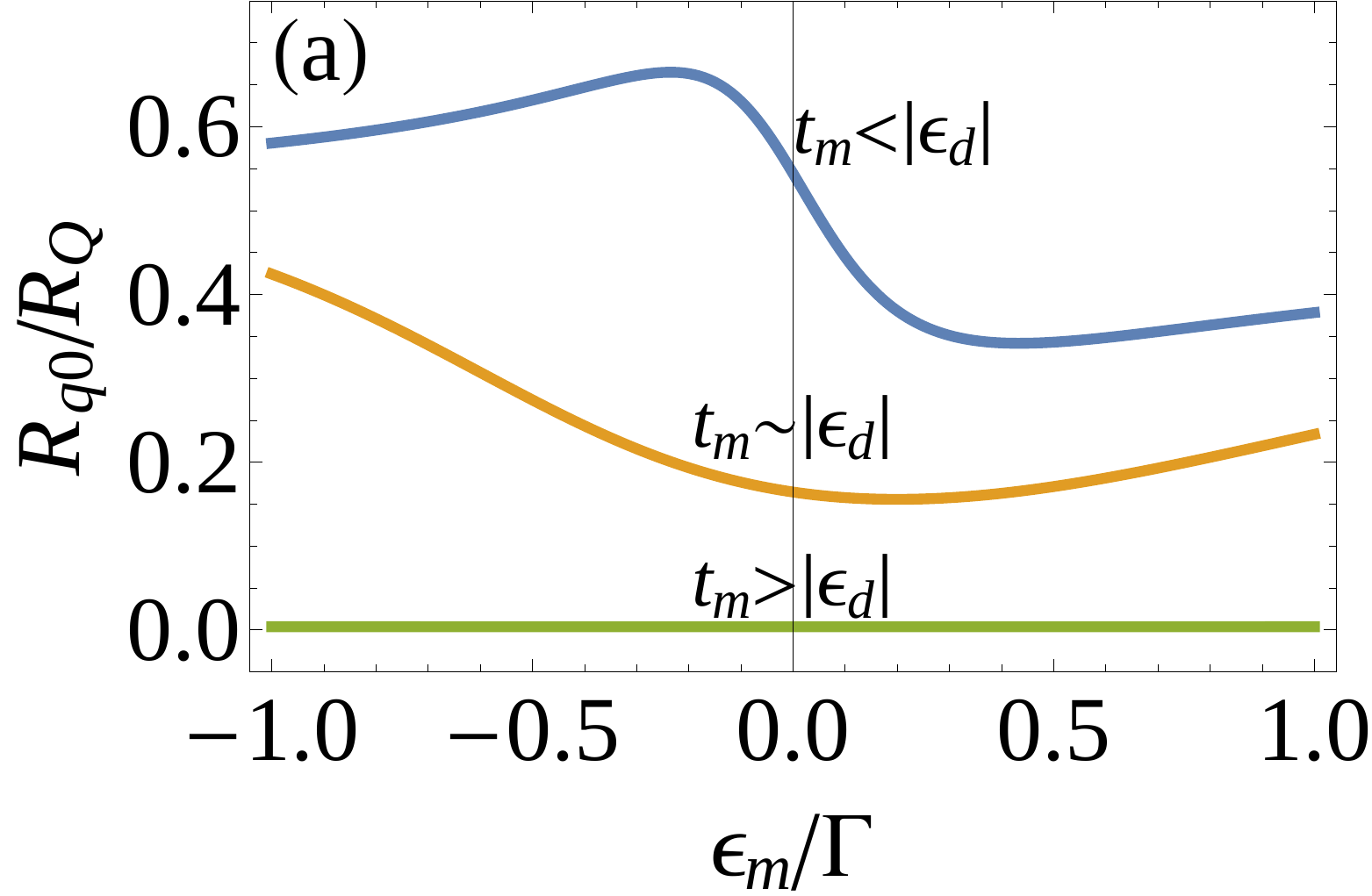}%
  \includegraphics[width=.5\textwidth]{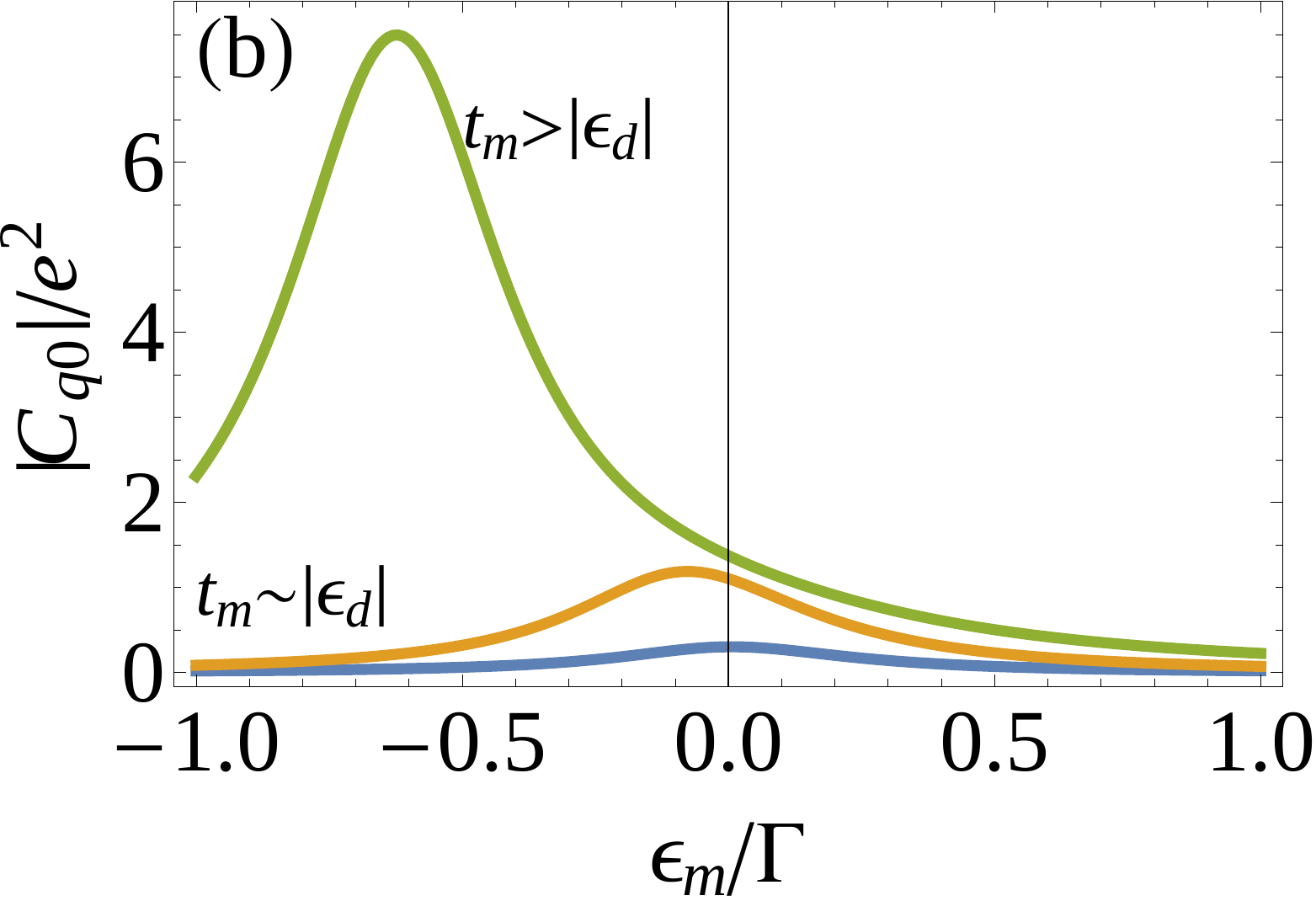}%
  \caption{(a) The low-frequency relaxation resistance $R_{q0}$ and (b) quantum
    capacitance $C_{q0}$ as functions of $\epsilon_m$ at zero temperature for
    the \Ni system in different regimes: $t_m<|\epsilon_d|$,
    $t_m\sim|\epsilon_d|$, and $t_m>|\epsilon_d|$ with
    $\epsilon_d/\Gamma = -0.1$.}
  \label{fig:N2}
\end{figure}

Now we examine the non-Majorana counterpart systems with finite $\epsilon_m$
for comparison.
First, since in the \Nii system the dot remains disconnected from the lead even
for finite $\epsilon_m$, no dissipation takes place, that is, $R_q=0$. It is
clearly distinguished from finite $R_{q0}$ of the Majorana system [see
\figref{fig:RqCq}].
Next, we consider the \Ni system whose $R_{q0}$ and $C_{q0}$ are shown in
\figref{fig:N2}. There are two clear differences when compared to the Majorana
system.
(1) Both $R_{q0}$ and $C_{q0}$ are asymmetric with respect to the change
$\epsilon_m\to-\epsilon_m$, while they are symmetric in the Majorana
system. The excitation energies $\pm\epsilon_{1,2}^M$ of the Majorana
system is clearly invariant under $\epsilon_m\to-\epsilon_m$, while
$\pm\epsilon_{1,2}^N$ are not. It is because the Majorana system is
inherently particle-hole symmetric.
(2) $R_{q0}$ decreases with increasing $t_m$ in the \Ni system [see
\figref{fig:N2}~(a)] while the Majorana system exhibits the opposite behavior
[see \figref{fig:RqCq}~(c)]. It is also attributed to the difference in the
excitation structure: For the Majorana case, $\delta\epsilon_1$ remains small
even for large $t_m$ [see \figref{fig:exc}~(b)], which indicates that the
energy cost for the p-h pair generation does not change so much. Instead, the
larger $t_m$ enhances the tunneling of electrons between the dot and the lead
so that the p-h generation in the lead is enhanced. On the other hand, in the
\Ni system, the excitation energy $\delta\epsilon_1$ increases with $t_m$
(though not monotonically) [see \figref{fig:exc}~(b)], increasing the energy
cost for the p-h pairs and suppressing its generation.

\section{Conclusion}
\label{set:conclusion}

We have investigated the effect of the Majorana bound states on the charging
and the dissipation of the quantum-dot system. It is found that the relaxation
resistance vanishes completely for $\epsilon_m=0$ and can be quite enhanced on
the dot resonance condition for $\epsilon_m\ne0$. The quantum capacitance is
found to follow the different dependence on systems parameters and temperature
when compared to that of non-Majorana systems. In order to identify clearly the
physical features related to the Majorana mode, we have considered two
non-Majorana systems. The \Nii system, having a localized bound states
permanently decoupled from the lead, does not dissipate, which is distinguished
from the Majorana case having finite dissipation for $\epsilon_m\ne0$. On the
other hand, the \Ni system always opens a dissipation channel and has
asymmetric dependence of the relaxation resistance and the quantum capacitance
on $\epsilon_m$ since it lacks the particle-hole symmetry which is inherent in
the Majorana case.  Therefore, the ac response of the Majorana RC circuit can
be used as another unambiguous method of detecting the elusive Majorana
fermions.

Similar ac responses such as the complete vanishing or
resonance-induced-enhancing of the relaxation resistance have been predicted in
the RC circuit composed of chiral Majorana edge modes around the
two-dimensional topological superconductor \cite{LeeMC2014aug}. It should be
noted that in our system the dissipation does not take place in the
superconductor while it does in the two-dimensional case. It means that the
Majorana-related features are quite common whether or not the Majorana modes
take part in the dissipation. In the experimental point of view, the
topological superconducting wire is much easier to implement than the
two-dimensional one. So we expect that our system is more adequate to observe
the non-trivial ac response of the Majorana modes.

\section{Acknowledgment}

This work was supported by the the National Research Foundation (Grant
Nos. 2011-0030046).

\section*{References}

\bibliographystyle{elsarticle-num}
\bibliography{paper}

\end{document}